\documentclass{new}
\usepackage{wrapfig}

\begin{document}

\vspace{20pt}

\markboth{Brodsky}{Applications of AdS/CFT Duality to QCD}

\title{Applications of AdS/CFT Duality to QCD}

\author{Stanley J. Brodsky}
\address{Stanford Linear Accelerator Center, Stanford University, Stanford, California 94309, USA}

\author{Guy F. de T\'eramond}
\address{Universidad de Costa Rica, San Jos\'e, Costa Rica}

\vfill

\maketitle

\begin{abstract}
Even though quantum chromodynamics is a broken conformal theory, the
AdS/CFT correspondence has led to important insights into the
properties of QCD.  For example, as shown by Polchinski and
Strassler, dimensional counting rules for the power-law falloff of
hadron scattering amplitudes  follow from dual holographic models
with conformal behavior at short distances and confinement at large
distances.  We find that one also obtains a remarkable
representation of the entire light-quark meson and baryon spectrum,
including all orbital excitations, based on only one mass parameter.
We also show how hadron light-front wavefunctions and hadron form
factors in both the space-like and time-like regions can be
predicted.
\end{abstract}

\section{Introduction}
The central mathematical principle underlying AdS/CFT duality  is
the fact that the group $SO(2,4)$ of Poincar\'e and conformal
transformations of physical $3+1$ space-time has an elegant
mathematical representation on ${\rm AdS}_5$ space where the fifth
dimension has the anti-de Sitter warped metric.  The group of
conformal transformations $SO(2,4)$ in 3+1 space  is isomorphic to
the group of isometries of AdS space, $x^\mu \to \lambda x^\mu$, $r
\to r/\lambda$, where $r$ represents the coordinate in the fifth
dimension. The dynamics at $x^2 \to 0$ in 3+1 space thus matches the
behavior of the theory at the boundary $r \to \infty.$ This allows one
to map the physics of quantum field theories with conformal symmetry
to an equivalent description in which scale transformations have an
explicit representation in AdS space.

Even though quantum chromodynamics  is a broken conformal theory,
the AdS/CFT correspondence  has led to important insights into the
properties of QCD. For example, as shown by  Polchinski and
Strassler,\cite{Polchinski:2001tt} the AdS/CFT duality, modified to
give a mass scale, provides a nonperturbative derivation of the
empirically successful dimensional counting
rules\cite{Brodsky:1973kr,Matveev:1973ra} for the leading power-law
fall-off of  the hard exclusive scattering amplitudes of the bound
states of the gauge theory. The modified theory generates the hard
behavior expected from QCD instead of the soft behavior
characteristic of strings. Other important applications include the
description  of spacelike hadron form factors at large transverse
momentum\cite{Polchinski:2001ju} and deep inelastic scattering
structure functions at small $x$.\cite{Polchinski:2002jw}    The
power falloff of hadronic light-front wave functions (LFWF)
including states with nonzero orbital angular momentum is also
predicted.\cite{Brodsky:2003px}

In the original formulation by Maldacena,\cite{Maldacena:1997re} a
correspondence was established between a  supergravity  string
theory on a curved background and a conformally invariant
$\mathcal{N} = 4$ super Yang-Mills theory in four-dimensional
space-time. The higher dimensional theory is $AdS_5 \times S^5$
where $R = ({4 \pi g_s N_C})^{1/4} \alpha_s'^{1/2}$ is the radius of
AdS and the radius of the five-sphere and $\alpha_s'^{1/2}$ is the
string scale. The extra dimensions of the five-dimensional sphere
$S^5$ correspond to the $SU(4) \sim SO(6)$ global symmetry which
rotates the particles present in the SYM supermultiplet in the
adjoint representation of $SU(N_C)$. In our application to QCD,
baryon number in QCD is represented as a Casimir constant on $S^5.$

The reason why AdS/CFT duality can have at least approximate
applicability to physical QCD is based on the fact that the
underlying classical QCD Lagrangian with massless quarks is
scale-invariant.\cite{Parisi:1972zy} One can thus take conformal
symmetry as an initial approximation to QCD, and then systematically
correct for its nonzero $\beta$ function and quark
masses.\cite{Brodsky:1985ve}  This ``conformal template"  approach
underlies the Banks-Zak method\cite{Banks:1981nn} for expansions of
QCD expressions near the conformal limit and the BLM
method\cite{Brodsky:1982gc} for setting the renormalization scale in
perturbative QCD applications. In the BLM method the corrections to
a perturbative series from the $\beta$-function are systematically
absorbed into the scale of the QCD running coupling. An important
example is the ``Generalized Crewther Relation"\cite{Brodsky:1995tb}
which relates the Bjorken and Gross-Llewellyn sum rules at the deep
inelastic scale $Q^2$ to the $e^+ e^-$ annihilation cross sections
at specific commensurate scales $ s^*(Q^2) \simeq 0.52~ Q^2$. The
Crewther relation\cite{Crewther:1972kn} was originally derived in
conformal theory; however, after BLM scale setting, it becomes a
fundamental test of physical QCD, with no uncertainties from the
choice of renormalization scale or scheme.

QCD is nearly conformal at large momentum transfers where asymptotic
freedom is applicable. Nevertheless, it is remarkable that
dimensional scaling for exclusive processes  is observed even at
relatively low momentum transfer where gluon exchanges involve
relatively soft momenta.\cite{deTeramond:2005kp}  The observed
scaling of hadron scattering amplitudes at moderate momentum
transfers can be understood if the QCD coupling has an infrared
fixed point.\cite{Brodsky:2004qb}   In this sense, QCD resembles a
strongly-coupled conformal theory.

\section{Hadron Spectra from AdS/CFT}

The duality between a gravity theory on $AdS_{d+1}$ space and a
conformal gauge theory at its $d$-dimensional boundary requires one
to match the partition functions at the $AdS$ boundary, $z = R^2/r
\to 0$. The physical string modes $\Phi(x,z) \sim e^{-i P \cdot x}
f(r)$, are plane waves along the Poincar\'e coordinates with
four-momentum  $P^\mu$ and hadronic invariant mass states $P_\mu
P^\mu = \mathcal{M}^2$. For large-$r$ or small-$z$, $f(r) \sim
r^{-\Delta}$, where the dimension $\Delta$ of the string mode must
be the same dimension as that of the interpolating operator
{\small$\mathcal{O}$} which creates a specific  hadron out of the
vacuum: $\langle P \vert \mathcal{O} \vert 0 \rangle  \neq 0$.

The physics of color confinement in QCD can be described in the
AdS/CFT approach by truncating the AdS space to the domain $r_0 < r
< \infty$  where $r_0 = \Lambda_{\rm QCD} R^2$.   The cutoff at
$r_0$ is dual to the introduction of a mass gap $\Lambda_{\rm QCD}$;
it breaks conformal invariance and is responsible for the generation
of a spectrum of color-singlet hadronic states. The truncation of
the AdS space insures that the distance between the colored quarks
and gluons as they stream into the fifth dimension is limited to $z
< z_0 = {1/\Lambda_{\rm QCD}}$. The resulting $3+1$ theory has both
color confinement at long distances and conformal behavior at short
distances.  The latter property allows one to derive dimensional
counting rules for form factors  and other hard exclusive processes
at high momentum transfer. This approach, which can be described as
a ``bottom-up" approach, has been successful in obtaining general
properties of the low-lying hadron spectra, chiral symmetry
breaking, and hadron couplings in AdS/QCD,\cite{Boschi-Filho:2002ta}
in addition to the hard scattering
predictions.\cite{Polchinski:2001tt,Polchinski:2002jw,Brodsky:2003px}

In this ``classical holographic model", the quarks and gluons
propagate into the truncated AdS interior according to the AdS
metric without interactions. In effect, their Wilson lines, which
are represented by open strings in the fifth dimension, are rigid.
The resulting equations for spin 0, $\frac{1}{2}$, 1 and
$\frac{3}{2}$ hadrons on $AdS_5 \times S^5$  lead to color-singlet
states with dimension $3, 4$  and $\frac{9}{2}$. Consequently, only
the hadronic states  (dimension-$3$) $J^P=0^-,1^-$ pseudoscalar and
vector mesons, the (dimension-$\frac{9}{2}$) $J^P=\frac{1}{2}^+,
\frac{3}{2}^+$ baryons, and the (dimension-$4$) $J^P= 0^+$ glueball
states, can be derived in the classical holographic
limit.\cite{deTeramond:2005su}  This description corresponds to the
valence Fock state as represented by the light-front Fock expansion.
Hadrons also fluctuate in particle number, in their color
representations ( such as the hidden-color
states\cite{Brodsky:1983vf} of the deuteron), as well as in internal
orbital angular momentum. The  higher Fock components of the hadrons
are manifestations of the quantum fluctuations of QCD;  these
correspond to the fluctuations of the bulk geometry about the fixed
AdS metric. Similarly, the orbital excitations of hadronic states
correspond to quantum fluctuations about the AdS
metric.\cite{Gubser:2002tv}  We thus can consistently identify
higher-spin hadrons with the fluctuations around the spin 0,
$\frac{1}{2}$, 1 and $\frac{3}{2}$ classical string solutions of the
$AdS_5$ sector.\cite{deTeramond:2005su}

\begin{figure}[tbh]
\centerline{\psfig{file=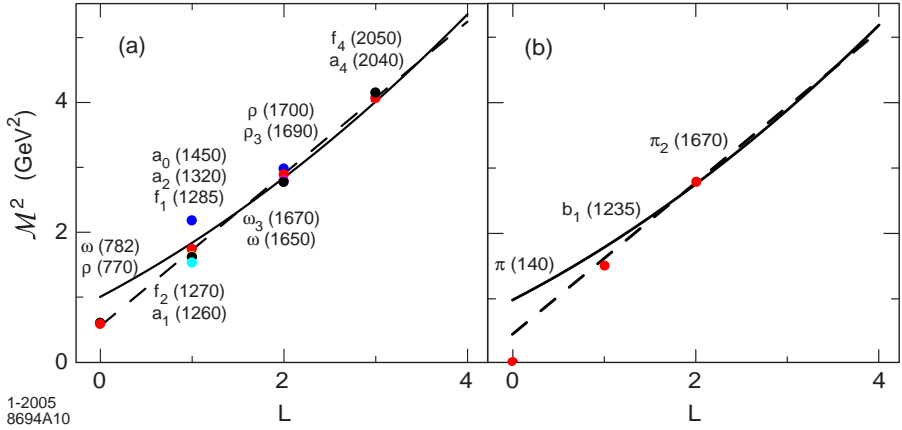,width=12.0cm}} \vspace*{8pt}
\caption{Light meson orbital states for $\Lambda_{\rm QCD} = 0.263$
GeV: (a) vector mesons and (b) pseudoscalar mesons. The dashed line
is a linear Regge trajectory with slope 1.16 ${\rm GeV}^2$.}
\label{fig:MesonSpec}
\end{figure}

\begin{figure}[tbh]
\centerline{\psfig{file=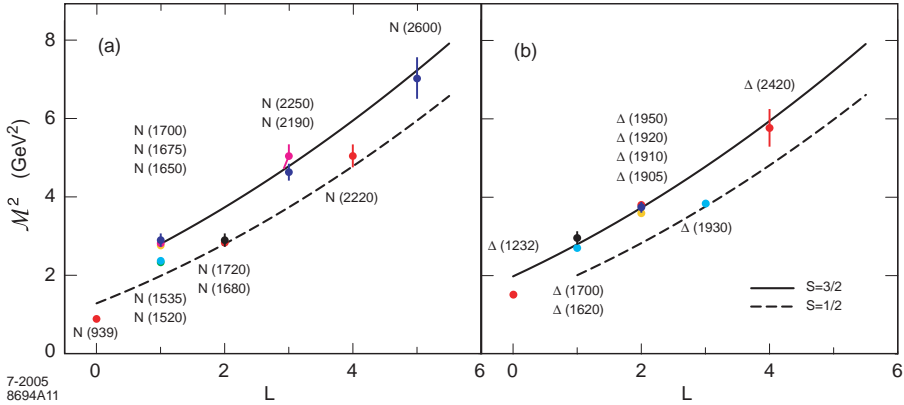,angle=0,width=12.0cm}}
\vspace*{8pt} \caption{Light baryon orbital spectrum for
 $\Lambda_{\rm QCD}$ = 0.22 GeV: (a) nucleons and (b) $\Delta$ states.}
\label{fig:BaryonSpec}
\end{figure}

As a specific example, consider the twist-two (dimension minus spin)
glueball interpolating operator $\mathcal{O}_{4 + L}^{\ell_1 \cdots
\ell_m} = F D_{\{\ell_1} \dots D_{\ell_m\}} F$ with total internal
space-time orbital momentum $L = \sum_{i=1}^m \ell_i$ and conformal
dimension $\Delta_L = 4 + L$. We match the large $r$ asymptotic
behavior of each string mode to the corresponding conformal
dimension of the boundary operators of each hadronic state while
maintaining conformal invariance. In the conformal limit, an $L$
quantum, which is identified with a quantum fluctuation about the
AdS geometry, corresponds to an effective five-dimensional mass
$\mu$ in the bulk side.  The allowed values of $\mu$ are uniquely
determined by requiring that asymptotically the dimensions become
spaced by integers, according to the spectral relation $(\mu R)^2 =
\Delta_L(\Delta_L - 4)$.\cite{deTeramond:2005su}  The
four-dimensional mass spectrum follows from the Dirichlet boundary
condition  $\Phi(x,z_o) = 0$, $z_0 = 1 / \Lambda_{\rm QCD}$, on the
AdS string amplitudes for  each  wave functions with  spin $<$ 2.
The eigen-spectrum is then determined from the zeros of Bessel
functions, $\beta_{\alpha,k}$. The predicted spectra
\cite{deTeramond:2005su}  of mesons and baryons with zero mass
quarks is shown in Figs.~\ref{fig:MesonSpec} and
\ref{fig:BaryonSpec}. The only parameter is $\Lambda_{\rm QCD} =
0.263$ GeV, and $0.22$ GeV for mesons and baryons, respectively.

\section{Dynamics from AdS/CFT}

\begin{figure}[htb]
\centerline{\psfig{file=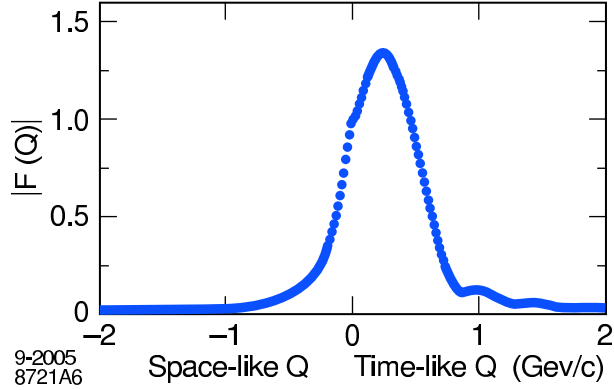,width=8.0cm}} \vspace*{8pt} \caption{Space-like and
time-like structure for the pion form factor in AdS/QCD.} \label{fig:FF}
\end{figure}

Current matrix elements in AdS/QCD are computed from the overlap of the
normalizable modes dual to the incoming
and outgoing hadron $\Phi_I$ and $\Phi_F$ and the
non-normalizable mode $J(x,z)$, dual to the external source
\begin{equation}
F(Q^2)_{I \to F}
\simeq R^{3 + 2\sigma} \int_0^{z_o} \frac{dz}{z^{3 + 2\sigma}}~
 \Phi_F(z)~J(Q,z)~\Phi_I(z),
\label{eq:FF}
\end{equation}
where $\sigma_n = \sum_{i=1}^n \sigma_i$ is the spin of the
interpolating operator $\mathcal{O}_n$, which creates an $n$-Fock
state $\vert n \rangle$ at the AdS boundary. $J(x,z)$  has the value
1 at zero momentum transfer, and as boundary limit the external
current, thus $A^\mu(x,z) = \epsilon^\mu e^{i Q \cdot x} J(Q,z)$.
The solution to the AdS wave equation subject to  boundary
conditions at  $Q = 0$ and $z \to 0$ is\cite{Polchinski:2002jw}
$J(Q,z) = z Q K_1(z Q)$. At large enough $Q \sim r/R^2$, the
important contribution to (\ref{eq:FF}) is from the region near $z
\sim 1/Q$. At small $z$, the $n$-mode $\Phi^{(n)}$ scales as
$\Phi^{(n)} \sim z^{\Delta_n}$, and we recover the power law
scaling,\cite{Brodsky:1973kr}  $F(Q^2) \to \left[1/Q^2\right]^{\tau
- 1}$, where the twist $\tau = \Delta_n - \sigma_n$, is equal to the
number of partons, $\tau_n = n$. A numerical computation for the
pion form factor gives the results shown in Fig.~\ref{fig:FF}, where
the resonant structure in the time-like region from the AdS cavity
modes is apparent.

\section{AdS/CFT Predictions for Light-Front Wavefunctions}

The AdS/QCD correspondence provides a simple description of hadrons
at the amplitude level by mapping string modes to the impact space
representation of LFWFs. It is useful to define the partonic
variables $ x_i \vec r_{\perp i} = \vec R_\perp + \vec b_{\perp i}$,
where $\vec r_{\perp i}$ are the physical position coordinates,
$\vec b_{\perp i}$  are frame-independent internal coordinates,
$\sum_i \vec b_{\perp i} = 0$,  and $\vec R_\perp$ is the hadron
transverse center of momentum $\vec R_\perp = \sum_i x_i \vec
r_{\perp i}$, $\sum_i x_i = 1$.   We find for a two-parton LFWF the
Lorentz-invariant form
\begin{equation}
\widetilde{\psi}_L(x, \vec b_{\perp}) = C ~x(1-x)
~\frac{J_{1+L}\left(\vert\vec b_\perp\vert
\sqrt{x(1-x)}~\beta_{1+L,k} \Lambda_{\rm QCD} \right)} {\vert\vec
b_\perp\vert \sqrt{x(1-x)}}. \label{eq:LFWFbM}
\end{equation}
The $ \beta_{1+L,k}$ are the zeroes of the Bessel functions
reflecting the Dirichlet boundary condition. The variable
$\zeta=\vert\vec b_\perp\vert \sqrt{x(1-x)},$ $0 \leq \zeta \leq
\Lambda^{-1}_{QCD}$, represents the invariant separation between
quarks. In the case of a two-parton state,  it gives a direct
relation between the scale of the invariant separation between
quarks, $\zeta$, and the holographic coordinate in AdS space: $\zeta
= z = R^2/r$. The ground state and first orbital eigenmode are
depicted in the figure below.
\begin{figure}[tbh]
\centerline{\psfig{file=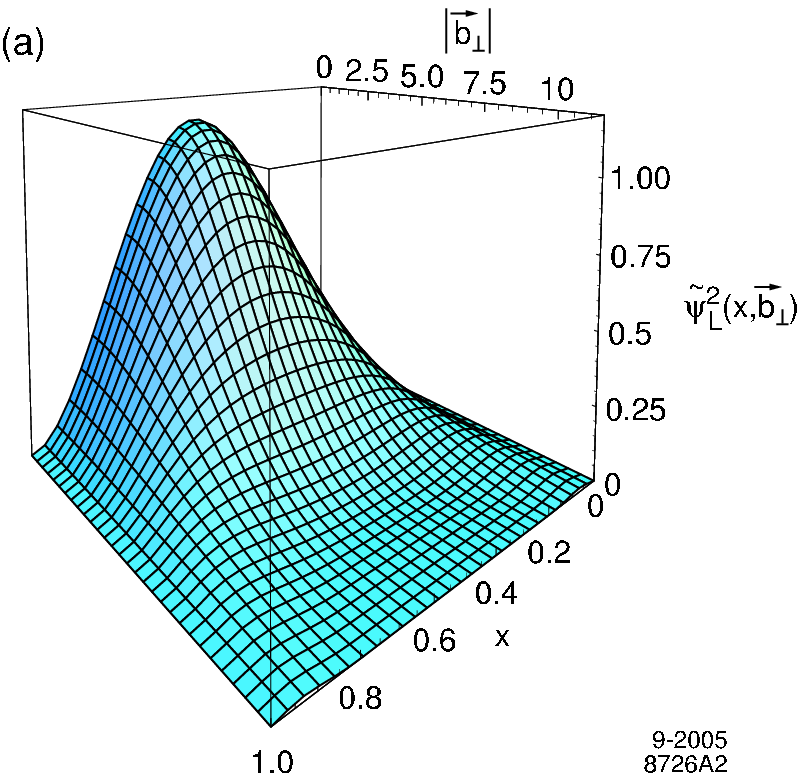,angle=0,width=6.0cm}
           {\psfig{file=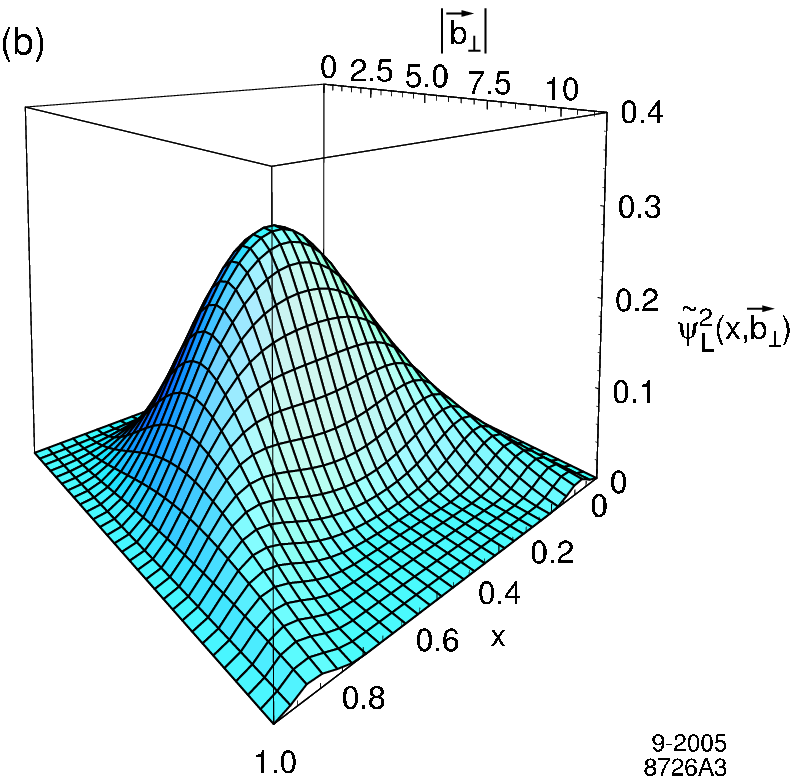,angle=0,width=6.0cm}}}
           \vspace*{8pt} \caption[*]{Prediction for
the square of the two-parton bound-state light-front wave function
$\widetilde{\psi}_L(x,\vec b_\perp)$ as function of the constituents
longitudinal momentum fraction $x$ and $1-x$ and the impact space
relative coordinate $\vec b_\perp$: (a) $L=0$ and (b) $L=1$.}
\label{fig:LFWFb}
\end{figure}

The holographic model is quite successful in describing the known
light hadron spectrum. The only mass  scale is $\Lambda_{QCD}$. The
model incorporates confinement and conformal symmetry. Only
dimension-$3, \frac{9}{2}$ and 4 states $\bar q q$, $q q q$, and  $g
g$ appear in the duality at the classical level. As we have
described, non-zero orbital angular momentum and higher Fock-states
require the introduction of quantum fluctuations. The model gives a
simple description of the structure of hadronic form factors and
LFWFs, which can be used as an initial approximation
to the actual eigensolutions of the light-front Hamiltonian for QCD.
It also explains the suppression of the odderon. The dominance of
the quark-interchange mechanism in hard exclusive processes also
emerges naturally from the classical duality of the holographic
model.

\section*{Acknowledgements}
Presented by SJB at
QCD 2005,
20 June 2005,
Beijing, China.
This work was supported by the Department
of Energy contract DE--AC02--76SF00515.

\end{document}